\documentclass{emulateapj}
\usepackage{natbib}
\usepackage{graphicx}
\usepackage{epstopdf}
\usepackage[english]{babel}
\usepackage[colorlinks,linkcolor={blue},citecolor={blue},urlcolor={blue}]{hyperref}
\def\mpchi{\,h^{-1}{\rm {Mpc}}}

\def\s{\mathbf{s}}
\def\be{\begin{equation}}
\def\ee{\end{equation}}
\def\ba{\begin{eqnarray}}
\def\ea{\end{eqnarray}}

\begin{document}

\title{Stellar mass and color dependence of the three-point correlation function of galaxies in the local universe}

\author{Hong Guo\altaffilmark{1,2,3},
Cheng Li\altaffilmark{1}, Y. P. Jing\altaffilmark{4}, Gerhard
B\"{o}rner\altaffilmark{5}}

\altaffiltext{1}{Partner Group of the Max Planck Institute for Astrophysics
at the Shanghai Astronomical Observatory and Key Laboratory for Research in
Galaxies and Cosmology of Chinese Academy of Sciences, Nandan Road 80,
Shanghai 200030, China} \altaffiltext{2}{Department of Astronomy, Case
Western Reserve University, OH 44106, USA} \altaffiltext{3}{Department of
Physics and Astronomy, University of Utah, 115 South 1400 East, Salt Lake
City, UT 84112, USA} \altaffiltext{4}{Center for Astronomy and Astrophysics,
Physics Department, Shanghai Jiao Tong University, Shanghai 200240, China}
\altaffiltext{5}{Max-Planck-Institut f\"{u}r Astrophysik,
Karl-Schwarzschild-Strasse 1, 85748 Garching, Germany}

\begin{abstract}
The three-point correlation function (3PCF) for galaxies provides an
opportunity to measure the non-Gaussianity generated from nonlinear structure
formation and also probes information about galaxy formation and evolution
that is generally not available from the two-point correlation function
(2PCF). We measure the 3PCF of the Sloan Digital Sky Survey DR7 main sample
galaxies in both redshift and projected spaces on scales up to $40\mpchi$. We
explore the dependence of the 3PCF on galaxy stellar mass and color in order
to constrain the formation and evolution for galaxies of different
properties. The study of the dependence on these properties also helps better
constrain the relation between galaxy stellar mass and color and the
properties of their hosting dark-matter halos. We focus on the study of the
reduced 3PCF, $Q$, defined as the ratio between the 3PCF and the sum of the
products of the 2PCFs. We find a very weak stellar mass dependence of $Q$ in
both redshift and projected spaces. On small scales, more massive galaxies
tend to have slightly higher amplitudes of $Q$. The shape dependence of $Q$
is also weak on these small scales, regardless of stellar mass and color. The
reduced 3PCF has a strong color dependence for the low-mass galaxies, while
no significant dependence on color is found for the high-mass galaxies.
Low-mass red galaxies have higher amplitudes and stronger shape dependence of
the reduced 3PCF than the blue galaxies, implying that these low-mass red
galaxies tend to populate filamentary structures. The linear galaxy bias
model fails to interpret the color dependence of $Q$, emphasizing the
importance of a nonvanishing quadratic bias parameter in the correct modeling
of the galaxy color distribution.
\end{abstract}


\section{Introduction}
In the current paradigm of galaxy formation within a merging hierarchy of
dark-matter halos, galaxies form when gas is able to cool, condense, and form
stars at the centers of dark-matter halos \citep{White-Rees-78}. Galaxy
formation thus involves complicated baryonic processes that play crucial
roles on galaxy and cluster scales but are poorly understood in many aspects.
One of the resulting effects is that the galaxy distribution is biased
relative to the dark matter distribution and such biasing depends strongly on
both spatial scale and galaxy properties \citep[see
e.g.,][]{Kaiser-84,Bardeen-86}. Studies of galaxy distribution are expected
to provide powerful constraints on models of galaxy formation and evolution.

It is common to use the two-point correlation function (2PCF), $\xi(r)$, to
quantify the galaxy clustering in both theories and galaxy surveys
\citep[]{Davis-Geller-76,Groth-Peebles-77,Davis-85,Davis-88,Hamilton-88,
White-Tully-Davis-88,Boerner-Mo-Zhou-89,Einasto-91,Park-94,Loveday-95,Benoist-96,
Guzzo-97,Beisbart-Kerscher-00,Norberg-01,Zehavi-02,Zehavi-05,Zehavi-11,Li-06a,
Li-07,Skibba-06,Wang-07,Zheng-Coil-Zehavi-07,Swanson-08,Watson-11,Guo-13}.
The 2PCF measures the excess probability of finding pairs in the universe. A
Gaussian-distributed density field can be fully described by the 2PCF, or its
Fourier-space counterpart, the power spectrum. However, non-Gaussianity
naturally arises in the nonlinear evolution of the density fluctuation
through the mode-coupling of different scales \citep[see
e.g.,][]{Bernardeau-02}. To quantify the non-Gaussianity and better
understand the galaxy distribution, we would require the next-order
statistics in the hierarchy of $N$-point correlation functions, the
three-point correlation function (3PCF), $\zeta(r_1,r_2,r_3)$, which can be
used to derive the nonlinear galaxy bias. A consequence of the additional
information provided by the 3PCF is that we can break the degeneracy of
galaxy bias and $\sigma_8$ (the rms of the matter density field in the
$8\mpchi$ sphere) in the 2PCF, which also helps constrain the cosmological
parameters
\citep{Gaztanaga-Frieman-94,Gaztanaga-05,Scoccimarro-01,Verde-02,Jing-Borner-04a,
Zheng-04,Gaztanaga-05,Pan-05,Nishimichi-07,
Ross-08,Ross-09,Guo-Jing-09b,Marin-11,McBride-11b,Marin-13,Pollack-13}.

For the study of the 3PCF, it is generally more convenient to introduce the
reduced 3PCF, $Q(r_1,r_2,r_3)$, defined as the ratio between the 3PCF and the
sum of the products of the 2PCFs,
\begin{equation}
Q(r_1,r_2,r_3)=\frac{\zeta(r_1,r_2,r_3)}{\xi(r_1)\xi(r_2)+\xi(r_2)\xi(r_3)+\xi(r_3)\xi(r_1)}
\end{equation}
Such a hierarchical scaling ($\zeta\propto\xi^2$) was introduced by
\cite{Groth-Peebles-77}, assuming $Q$ to be a constant. Although subsequent
theories and observations found that $Q$ is generally dependent on the size
and shape of the triangle ($r_1,r_2,r_3$) \citep[see e.g.,][]{Bernardeau-02},
it is still more natural to use $Q$ as the measurement of non-Gaussianity
generated from gravitational clustering. In observation, the redshift-space
distortion (RSD) effect will significantly change the shape and amplitude of
both the 2PCF and the 3PCF, because the galaxy peculiar velocities prevent us
from measuring the real radial distribution of the galaxies. The projected
correlation function is then used to minimize the influence of the RSD. Since
the projected correlation function can be easily converted to the real-space
correlation function \citep{Davis-85}, the projected 3PCF was proposed to
represent the actual 3PCF in real space. \cite{Jing-Borner-98} first applied
this statistic to the Las Campanas Redshift Survey.

In order to use the 3PCF to constrain the galaxy formation and evolution
models, knowledge of the relation between galaxies and dark-matter halos is
important. The clustering dependence of the 2PCF on galaxy properties is
extensively investigated in the literature to infer how galaxies of different
properties populate the dark-matter halos of different masses, which is
commonly referred to as the halo occupation distribution (HOD)
\citep[e.g.,][]{Zehavi-05,Zehavi-11,Zheng-07,Zheng-09,Coupon-12}. However,
the 2PCF itself is not enough to break the degeneracy in the theoretical
model parameters. The measurements of the clustering dependence on galaxy
properties from the 3PCF will further help put a strong constraint on the
allowed HOD parameters \citep{Kulkarni-07}.

\cite{Jing-Borner-04a} used the Two-degree Field Galaxy Redshift Survey
\citep[2dFGRS;][]{Colless-01} to measure the luminosity dependence of the
reduced 3PCF, $Q(r_1,r_2,r_3)$, in both redshift and projected spaces. They
found a small but significant trend: more luminous galaxies have lower
amplitudes of $Q$. \cite{Gaztanaga-05} also measured the redshift-space $Q$
with 2dFGRS and found a weak tendency for $Q$ to decrease with increasing
luminosity in the nonlinear regime. \cite{Kayo-04} and \cite{Nichol-06}
employed the early data release of the Sloan Digital Sky Survey
\citep[SDSS;][]{York-00} and found no significant dependence of $Q$ on galaxy
morphology, color, or luminosity. Considering the large measurement errors,
their results are statistically consistent with \cite{Jing-Borner-04a} and
\cite{Gaztanaga-05}. \cite{McBride-11a} explored the dependence of $Q$ on
luminosity and color in redshift and projected spaces using SDSS Data Release
6. They found a similar luminosity and color dependence to previous work and
inferred that the weak shape dependence of $Q$ on small scales is caused by
the RSD.

Apart from the analysis of the observational data, another way to study the
effect of the RSD on the 3PCF is to use numerical simulations
\citep[e.g.,][]{Gaztanaga-Scoccimarro-05,Marin-08}. \cite{Marin-08} used
$N$-body simulations to construct mock galaxy catalogs and compared $Q$ in
both real and redshift spaces. They found the same luminosity dependence of
$Q$ as in the observations and they argued that the color dependence of $Q$
appears stronger than the luminosity dependence, which is consistent with the
finding of \cite{Gaztanaga-05}.

In this paper, we will focus on measuring the dependence of the 3PCF on the
galaxy stellar mass and the dependence on color in certain stellar mass bins.
We use the final release data \citep[DR7;][]{Abazajian-09} of SDSS to measure
the 3PCF in both redshift and projected spaces in the local universe. Since
the stellar mass dependence of the 3PCF has never been studied before, it
will provide powerful constraints on the connection between galaxy stellar
mass and their hosting dark matter halo mass, which will be explored in our
future work. For consistency with previous work, we also briefly investigate
the luminosity dependence of the 3PCF in our sample. We extend the previous
analysis by exploring galaxy samples in a broader range of scales and
triangle configurations, in order to clarify whether the weak shape
dependence of $Q$ is really caused by the RSD.

The paper is constructed as follows: In Section~\ref{sec:data}, we briefly
describe the galaxy and mock samples. The methods of measuring the 2PCF and
3PCF are presented in Section~\ref{sec:methods}. We present our results of
the stellar mass and color dependence of the 3PCF in
Section~\ref{sec:results}. We summarize our results in
Section~\ref{sec:discussion}.

Throughout this paper we assume a cosmology model with a density parameter
$\Omega_m=0.3$, a cosmological constant $\Omega_\Lambda=0.7$, and a Hubble
constant $H_0=100h$kms$^{-1}$Mpc$^{-1}$ with $h=1.0$.

\section{Data}
\label{sec:data}

\subsection{SDSS Galaxy Samples}

The galaxy sample used in this paper is a magnitude-limited sample taken from
the New York University Value-Added Galaxy Catalog (NYU-VAGC), constructed by
\citet{Blanton-05c} based on SDSS DR7. This sample has formed the basis for
the recent studies on the galaxy distribution in the low-redshift universe
\citep[e.g.,][]{Li-12a, Li-12b, Li-Wang-Jing-13}. The sample contains about
half a million galaxies located in the main contiguous area of the survey in
the Northern Galactic Cap, with $r<17.6$, $-24<M_r<-16$, and
spectroscopically measured redshifts in the range of $0.001<z<0.5$. Here $r$
is the $r$-band Petrosian apparent magnitude, corrected for Galactic
extinction, and $M_r$ is the $r$-band Petrosian absolute magnitude, corrected
for evolution and $K$-corrected to $z=0.1$. The apparent magnitude limit is
chosen to select a sample that is uniform and complete over the entire area
of the survey \citep[see][]{Tegmark-04}. The median redshift of this sample
is $z=0.088$, with $10\%$ of the galaxies below $z=0.033$ and $10\%$ above
$z=0.16$.

The stellar mass of each galaxy in our sample accompanies the NYU-VAGC
release. It is estimated based on the redshift and the five-band {\tt
Petrosian} magnitudes from the SDSS photometric data, as described in detail
in \citet{Blanton-Roweis-07}. This estimate corrects implicitly for dust and
assumes a universal stellar initial mass function (IMF) of
\citet{Chabrier-03} form. As demonstrated in Appendix A of
\citet{Li-White-09}, once all estimates are adapted to assume the same IMF,
the estimated stellar masses agree quite well with those obtained from the
simple, single-color estimator of \citet{Bell-03} and also with those derived
by \citet{Kauffmann-03a} from a combination of SDSS photometry and
spectroscopy.

\begin{figure*}
\epsscale{1.0} \plotone{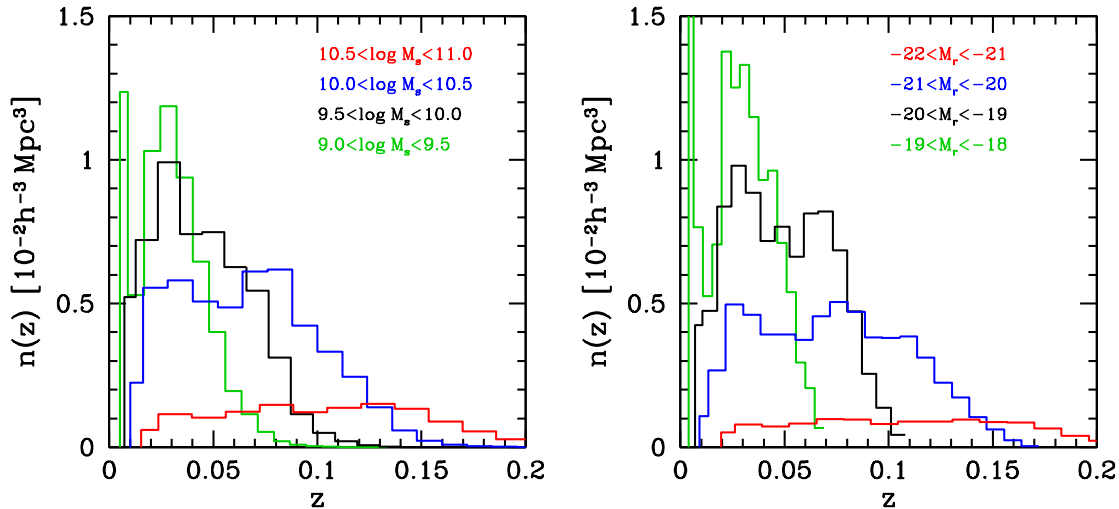} \caption{Number density distribution for
stellar mass (left panel) and luminosity subsamples (right panel). (A color
version of this figure is available in the online journal.)} \label{fig:dndz}
\end{figure*}

From this sample we select two sets of samples according to either $r$-band
absolute magnitude ($M_r$) or stellar mass ($\log M_s$). In each of these
samples, we further divide the galaxies into {\em red} and {\em blue}
populations according to their $g-r$ colors, following the
luminosity-dependent color cut determined in \citet{Li-06a}. Details of the
luminosity- and mass-selected samples, as well as their red/blue samples, are
listed in Table 1. The effective volume of each sample is calculated from the
summation \citep[e.g.,][]{Percival-10}
\begin{equation}
V_{\rm eff}=\sum_i\left(\frac{\bar{n}(z_i)P_0}{1+\bar{n}(z_i)P_0}\right)^2\Delta V(z_i),
\end{equation}
where $\bar{n}(z_i)$ is the mean number density in the volume shell $\Delta
V(z_i)$ at $z_i$ and $P_0=10^4h^{-3}{\rm Mpc}^3$. We show the number density
distribution, $n(z)$, for the different stellar-mass and luminosity
subsamples in Figure \ref{fig:dndz}. As expected, more massive or luminous
galaxies are spread over a larger volume. The least massive or the faintest
subsample only cover a small local volume, and their correlation function
measurements may suffer from a significant cosmic variance effect, as will be
shown in the following sections.
\begin{deluxetable*}{lccccccc} \tablewidth{0pt}
\tablecaption{\label{tab:sample}Flux-limited Samples of Different Luminosity
and Stellar Mass}
 \tablecomments{The stellar mass $M_s$ is in units of $h^{2}M_\odot$, where we assume
 $h\equiv1$ throughout the paper. The effective volume $V_{\rm eff}$ is in units of $h^{-3}{\rm Gpc}^3$.}
\startdata \hline Sample&$\log(M_s)$&$\bar{z}$&$V_{\rm eff}$&Number of
Galaxies&Red Galaxies&Blue Galaxies\\ \hline
M1& [$9.0,9.5$) & 0.046 & 0.011 & 32573 & 6658 & 25915\\
M2& [$9.5,10$) & 0.066 & 0.028 & 84024  & 30700 & 53324\\
M3& [$10,10.5$) & 0.095 & 0.069 & 181865 & 97918 & 83947 \\
M4& [$10.5,11$) & 0.137 & 0.129 & 143235 & 99932 & 43303\\
\hline
Sample&$M_r$&$\bar{z}$&$V_{\rm eff}$&Number of Galaxies&Red Galaxies&Blue Galaxies\\
\hline
L1& [$-19,-18$) & 0.042 & 0.006 & 36223  & 13127 & 23096\\
L2& [$-20,-19$) & 0.067 & 0.022 & 107241 & 52169 & 55072\\
L3& [$-21,-20$) & 0.102 & 0.070 & 189880 & 102128 & 87752\\
L4& [$-22,-21$) & 0.147 & 0.136 & 118728 & 73002 & 45726
\end{deluxetable*}

\subsection{Random Samples}

To obtain reliable estimates of the auto-correlation functions, each observed
sample in Table 1 must be compared with a ``random sample'' that is
unclustered but fills the same region of the sky and has the same,
position-dependent spectroscopic completeness and luminosity-dependent
redshift distribution. We construct our random samples from the observed
samples themselves, as described in detail in \citet{Li-06a}. For each real
galaxy, we generate $10$ sky positions at random within the mask of the
survey \citep[see][for details]{Blanton-05c}, and we assign to each of them
the properties of the real galaxy-- in particular, its values of redshift,
luminosity, and stellar mass as well as the position-dependent spectroscopic
completeness. This method is valid only for wide-field surveys like SDSS,
where randomizing sky positions should be sufficient to break the coherence
of the large-scale structures in the real sample. Extensive tests show that
random samples constructed in this way produce indistinguishable results from
those using the traditional method, which relies on the galaxy luminosity
function obtained from the same survey to determine the redshift-dependent
average number density of galaxies required for generating the random points
\citep{Li-06c}.

\subsection{Mock Galaxy Samples}

We construct a set of 80 mock SDSS galaxy catalogs from the Millennium
Simulation \citep{Springel-05a} using both the sky mask and the magnitude and
redshift limits of our real SDSS sample. The Millennium Simulation uses
$10^{10}$ particles to follow the dark-matter distribution in a cubic region
with 500$h^{-1}$Mpc on a side, assuming the concordance $\Lambda$ cold
dark-matter cosmology. Galaxy formation within the evolving dark-matter halos
is simulated in postprocessing using semi-analytic methods tuned to give a
good representation of the observed low-redshift galaxy population. Our mock
catalogs are based on the galaxy formation model of \citet{Croton-06} and are
constructed from the publicly available data at $z=0$ using the methodology
of \citet{Li-06c} and \citet{Li-07}. These mock catalogs allow us to derive
reasonable error estimates for the correlation functions we measure,
including both sampling and cosmic-variance uncertainties. We use all 80 mock
catalogs for estimating the errors of our 2PCF measurements. For the 3PCFs,
we use only 10 of the 80 mock catalogs randomly selected from the whole set,
in order to save computational time while obtaining a reasonably good
estimation of the errors. The volume of the Millennium simulation is
$0.125h^{-3}{\rm Gpc}^3$, which is enough for most of our stellar mass and
luminosity subsamples, and only slightly smaller than those of the samples
$M4$ and $L4$ (see Table \ref{tab:sample}). We note that the error estimation
in this paper makes use of only the diagonal elements of the covariance
matrix of the mock catalogues, and so the errors on different scales for a
given sample are correlated to varying degrees. However, we argue that this
is not an issue in our case where we use these errors mainly to indicate the
relative differences in the 2PCF and the 3PCF measured at given scale between
different samples. We also note that the possible systematic errors in our
estimation of the galaxy stellar mass and also the photometric errors in the
galaxy color are not considered. These errors would not change our
conclusions of the dependence on galaxy properties since they would not
significantly mix the galaxy samples.

\section{Correlation function measurements}
\label{sec:methods}

\subsection{Two-point Correlation Function}
We begin by estimating the two-dimensional, redshift-space, two-point
auto-correlation function, $\xi(r_p,r_\pi)$, for each of the samples listed
in Table \ref{tab:sample} using the estimator of \citet{Landy-Szalay-93}:
\begin{equation}
\xi(r_p,r_\pi)=\frac{\rm{DD-2DR+RR}}{\rm{RR}},
\label{eqn:landy-szalay}
\end{equation}
where $\rm{DD}$, $\rm{DR}$, and $\rm{RR}$ are the data-data, data-random and
random-random pair counts, and $r_p$ and $r_\pi$ are the pair separations
perpendicular and parallel to the line of sight. To normalize appropriately,
$\rm{RR}$ needs to be multiplied by $(N_g/N_r)^2$ and $\rm{DR}$ by $N_g/N_r$,
where $N_g$ and $N_r$ are the numbers of galaxies in the real and random
samples, respectively. In our case, $N_r=10\times N_g$. We have corrected the
effect of fiber collisions in the same way as in \citet{Li-06c} and
\citet{Li-07}. Briefly, we up-weight each galaxy pair separated by angular
distance $\theta$ by the ratio
$F(\theta)=[1+w_{pz}(\theta)]/[1+w_{sz}(\theta)]$, where $w_{sz}(\theta)$ and
$w_{pz}(\theta)$ are the angular 2PCFs of the spectroscopic and the parent
photometric samples, respectively. Detailed tests of the method can be found
in \citet{Li-06c} and \citet{Guo-12}.

Next, we integrate these two-dimensional correlation estimates over the
line-of-sight separation $r_\pi$ to obtain the projected auto-correlation
function, $w_p(r_p)$, as follows,
\begin{equation}
w_p(r_p) =
\int_{-r_{\pi,\rm{max}}}^{+r_{\pi,\rm{max}}} \xi(r_p,r_\pi)d r_\pi =
\sum_{i}\xi(r_p,r_\pi)\Delta r_{\pi,i},
\end{equation}
where we choose $r_{\pi,\rm{max}}=40h^{-1}$Mpc as the outer limit for the
integration depth (in order to reduce the noise from distant uncorrelated
regions) so that the summation for computing $w_p(r_p)$ runs from
$r_{\pi,1}=-39.5h^{-1}$Mpc to $r_{\pi,80}=39.5h^{-1}$Mpc, given that we use
bins of width $\Delta r_{\pi,i}=1h^{-1}$Mpc.

\begin{figure*}
\epsscale{1.0} \plotone{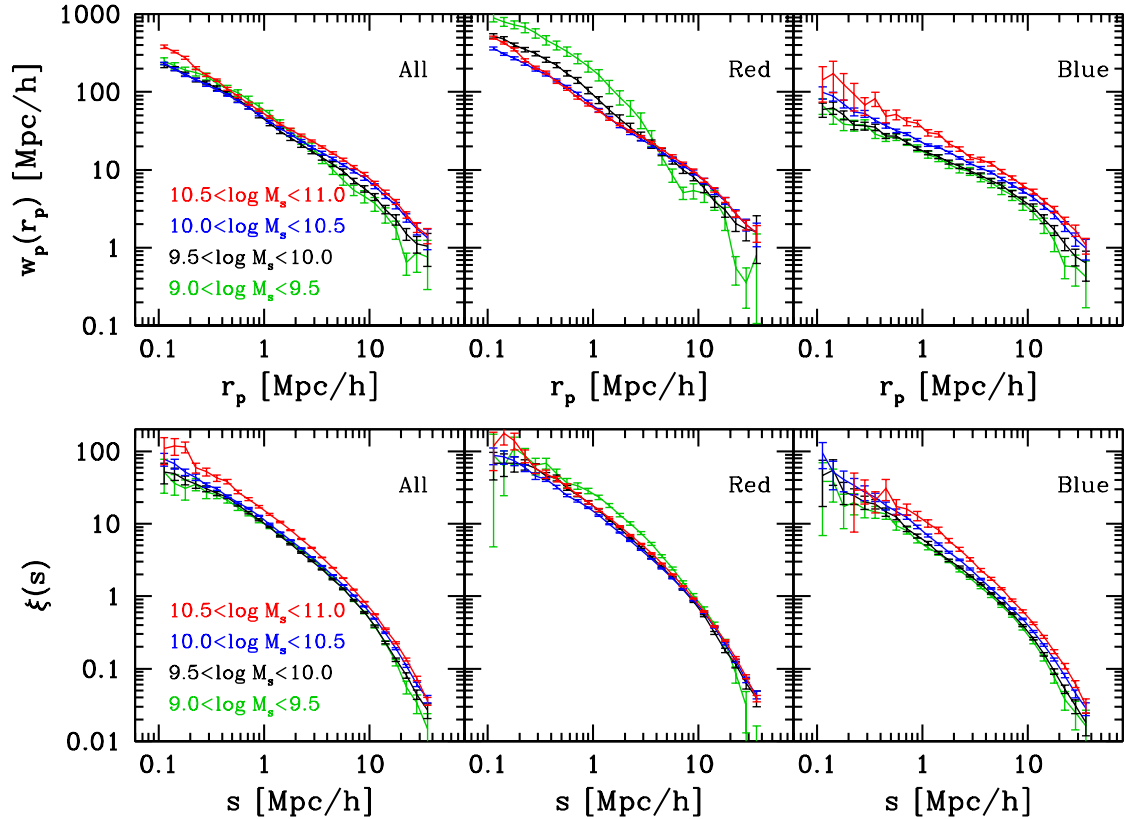} \caption{Projected- (top panels) and
redshift-space (bottom panels) 2PCFs for different stellar-mass samples.
Lines of different color denote different stellar samples. The measurements
of all the galaxies in each sample are shown in the left panels. The
corresponding correlation functions for the red and blue subsamples are also
shown in the middle and right panels, respectively. (A color version of this
figure is available in the online journal.)} \label{fig:2pcf}
\end{figure*}
In addition to $w_p(r_p)$, we also measure the one-dimensional
auto-correlation function in redshift space, $\xi(s)$, using the same
estimator as in Equation (\ref{eqn:landy-szalay}) but as a function of $s$,
the redshift-space three-dimensional (3D) pair separation, which is given by
$s^2=r_p^2+r_\pi^2$.

In Figure~\ref{fig:2pcf}, we present the measurements of $w_p(r_p)$ (top
panels) and $\xi(s)$ (bottom panels) for the four stellar-mass-selected
samples and also the color subsamples. The errors are estimated from the 80
mock catalogs as described in the previous section. We note that the
stellar-mass sample of $9<\log M_s<9.5$ shows a significantly different trend
from all other samples. It is caused by the large cosmic variance due to its
small volume, which can also be inferred from its small mean redshift,
$\bar{z}=0.046$. Similar behavior is also found for faint galaxies in the
luminosity bin of $-19<M_r<-18$ \citep{Zehavi-11}, which has a similar mean
redshift as this sample. Therefore, in order to better constrain the
dependence of the reduced 3PCFs on the stellar mass, we will ignore this
sample in the following sections. In all other cases, both $w_p(r_p)$ and
$\xi(s)$ show systematic trends with galaxy stellar mass, reflecting the
different masses of the hosting halos, with more massive galaxies residing in
more massive halos \citep[e.g.][]{Wang-Jing-10}. This trend is shown more
clearly in the blue subsamples. Because most of the blue galaxies of these
stellar masses are central galaxies, the different hosting-halo masses will
have stronger effects on the amplitudes of $w_p(r_p)$. On the contrary, the
two less massive red-galaxy samples ($\log M_s<10.0$) show strong small-scale
clustering. It can be caused by their high satellite fractions that increase
the clustering amplitude, though they are located in lower-mass halos
\citep{Li-07, Zehavi-11}.

The contribution to the projected 2PCF can be decomposed into two components:
a steeper inner part at pair separations below $\sim1 h^{-1}$Mpc and a
flatter outer part at larger separations. In the language of the ``halo
model'' \citep[see, e.g.,][]{Cooray-Sheth-02}, the inner part is called the
{\em one-halo} term, where the pair counts are mostly galaxy pairs in the
same halo, and the outer part is referred to as the {\em two-halo} term,
where galaxy pairs are mostly from two separate halos. The separation, where
the transition between the two terms occurs, increases with increasing
stellar mass. In the two-halo term, the amplitude of $w_p(r_p)$ is an
increasing function of stellar mass, while its slope shows weak or no stellar
mass dependence. This is consistent with the picture that, on large scales,
the bias in the galaxy distribution is related in a simple way to the bias in
the distribution of dark matter halos. When compared to the projected
correlation function, the $\xi(s)$ measurements are relatively suppressed on
small scales, leading to almost no transition between the one-halo and
two-halo terms. This is caused by the RSD effect on small scales and the
global infall toward high-density regions on large scales.

Both the projected and redshift-space 2PCFs have been extensively studied in
the literature using both SDSS-based samples \citep[see e.g.,][]{Zehavi-05,
Zehavi-11, Li-06a, Guo-13} and samples from other surveys. Our measurements
are well consistent with those previous determinations. In the current paper,
these two-point autocorrelation measurements for different stellar mass and
color bins will be used in what follows to determine the reduced 3PCFs.

\begin{figure*}
\epsscale{1.0} \plotone{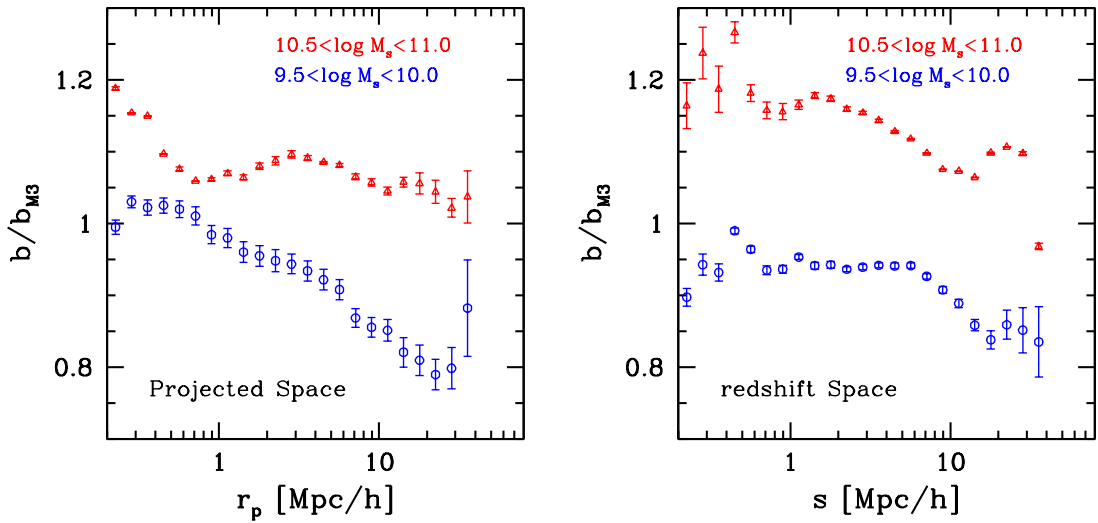} \caption{Measurements of relative
galaxy bias, $b/b_{M3}$, (Equation (\ref{eqn:rbias})) for the different
stellar mass subsamples, with triangles for $10.5<\log M_s<11$ and circles
for $9.5<\log M_s<10$. The measurements for the projected and redshift space
are shown in the left and right panels, respectively. (A color version of
this figure is available in the online journal.)} \label{fig:2pcfbias}
\end{figure*}
To quantify the galaxy bias of the different stellar-mass samples, we measure
the relative bias of the stellar-mass samples to the sample M3 as
\begin{equation}
b/b_{M3}=\sqrt{\xi(r)/\xi_{M3}(r)}, \label{eqn:rbias}
\end{equation}
where the 2PCF, $\xi(r)$, can be the redshift-space 2PCF $\xi(s)$ or the
projected-space 2PCF $w_p(r_p)$. We show in Figure \ref{fig:2pcfbias} the
measurements of the relative bias for the projected -(left panel) and
redshift-space (right panel) 2PCFs. Different symbols represent different
stellar-mass subsamples, with triangles for $10.5<\log M_s<11$ and circles
for $9.5<\log M_s<10$. As shown, the relative bias is strongly scale
dependent for scales less than $20\mpchi$, which reflects the different
spatial distributions of the stellar mass subsamples on these nonlinear
scales.

\subsection{Three-point Correlation Function}
By analogy with the case of the 2PCF, the definition of the 3PCF in the
projected space requires five free parameters, $r_{p12}$, $r_{p23}$,
$r_{p31}$, $r_{\pi12}$, and $r_{\pi23}$), where $r_{pij}$ and $r_{\pi ij}$
are the separations of objects $i$ and $j$ perpendicular to and along the
line of sight. The physical explanation of the 3PCF
$\zeta(r_{p12},r_{p23},r_{p31},r_{\pi12},r_{\pi23})$ is the excess
probability of finding three objects simultaneously in three volume elements
$dV_1$, $dV_2$, and $dV_3$ at positions $\s_1$, $\s_2$ and $\s_3$, excluding
the contributions from the 2PCFs,
\begin{eqnarray}
dp_{123}&=&\bar{n}(\s_1)\bar{n}(\s_2)\bar{n}(\s_3)\nonumber\\
&&\times[1+\xi(r_{p12},r_{\pi12})+\xi(r_{p23},r_{\pi23})+\xi(r_{p31},r_{\pi31})\nonumber\\
&&\quad+\zeta(r_{p12},r_{p23},r_{p31},r_{\pi12},r_{\pi23})]dV_1dV_2dV_3.
\end{eqnarray}
The projected 3PCF $\Pi(r_{p12},r_{p23},r_{p31})$ and its reduced counterpart
$Q_p(r_{p12},r_{p23},r_{p31})$ are defined as
\citep{Jing-Borner-98,Jing-Borner-04a}
\ba
&&\Pi(r_{p12},r_{p23},r_{p31})=\nonumber\\
&&\quad\int\,\zeta(r_{p12},r_{p23},r_{p31},r_{\pi12},r_{\pi23})dr_{\pi12}dr_{\pi23}\label{eqn:pi}\\
&&Q_p(r_{p12},r_{p23},r_{p31})=\frac{\Pi(r_{p12},r_{p23},r_{p31})}{w_p(r_{p12})w_p(r_{p23})+\mbox{cyc}}.
\ea
Similarly, the redshift-space reduced 3PCF is defined by
\begin{equation}
Q_s(s_{12},s_{23},s_{31})=\frac{\zeta(s_{12},s_{23},s_{31})}{\xi(s_{12})\xi(s_{23})+\mbox{cyc}}\quad.
\label{eqn:qs}
\end{equation}

We use the following estimator to measure $\zeta$ and $\Pi$ in our samples
\citep{Szapudi-Szalay-98},
\begin{equation}
\zeta=\frac{\rm{DDD-3DDR+3DRR-RRR}}{\rm{RRR}},\label{eqn:estimator}
\end{equation}
where $\rm{DDD}$, $\rm{DDR}$, $\rm{DRR}$, and $\rm{RRR}$ are the normalized
number counts of triplets in each bin measured from the data (represented by
$\rm{D}$) and random samples (represented by $\rm{R}$). We correct the
fiber-collision effect by adding the weight of each fiber-collided galaxy to
its nearest neighbor, which should produce the correct clustering amplitude
beyond the fiber-collision scale \citep[${\sim}0.1\mpchi$ at
$z{\sim}0.1$;][]{Guo-12}.

Instead of using $(r_{1},r_{2},r_{3})$ to represent the triangle, we use the
parameterization of $(r_{1},r_{2},\theta)$, with $r_{2}\ge r_{1}$ and
$\theta=\arccos[(r_1^2+r_2^2-r_3^2)/2r_1r_2]$. We note that in this
parameterization, one triangle is counted three times in different bins of
$(r_{1},r_{2},\theta)$, but that does not affect our measurements of the
configuration dependence. Following \cite{Jing-Borner-04a}, we use a linear
binning scheme for $\theta$. The choice of $\Delta\theta$ is
$\Delta\theta=\pi/20$, which is small enough to trace the dependence on
$\theta$. Unless a very large bin size (e.g., $\Delta\theta=\pi/5$) is used,
decreasing the bin size of $\theta$ in our samples does not improve the
results but only reduces the signal-to-noise ratio \citep[S/N;][]{Marin-11}.

As in the 2PCF, we use the same logarithmic binning schemes for $r_1$ and
$r_2$, i.e. $\Delta\log r_1=\Delta\log r_2$. In order to have a reasonable
S/N, we use a different $\Delta\log r_1$ on different scales, but the
comparisons of the reduced 3PCF, $Q$, for galaxies of different properties
are made under the same binning scheme. We note that the amplitude and shape
of $Q$ is coupled with the choice of bin sizes for $r_1$ and $r_2$, because
the reduced 3PCF, $Q$, is generally dependent on the scale. The range of
$r_3$ for the same angular separation $\theta$ would also vary with
$\Delta\log r_1$. For example, for certain $r_1$ and $r_2$, if a very large
$\Delta\log r_1$ is adopted, the values of $Q(r_1,r_2,\theta)$ are then
effectively smoothing over a large range of scales, which may change its
shape dependence. So in this paper, we use a small-as-possible $\Delta\log
r_1$ when a reasonable S/N is achieved.

Another commonly used parameterization of $Q$ is $(r, u,\theta)$, where
$r\equiv r_1$, $u\equiv r_2/r_1$, and $\theta$ is the angle between $r_1$ and
$r_2$
\citep[e.g.,][]{Scoccimarro-01,Kayo-04,Nichol-06,Kulkarni-07,Guo-Jing-09a}.
The definition is similar to ours, except for the replacement of $r_2$ by
$u$. Such a parameterization is introduced to decompose the elements of the
triangle into scale (denoted by $r$) and shape (represented by $u$ and
$\theta$) components. We note that in this parameterization, $\Delta\log
r_2=\Delta\log r_1+\Delta\log u$, which means that the bin width of $r_2$ is
always larger than that of $r_1$. Including $u$ complicates the analysis,
because we also need to take into account the bin width of $u$. In order to
have a small enough bin width, we would rather use $r_2$ instead of $u$ and
require that $\Delta\log r_2=\Delta\log r_1$, which cannot be achieved with
the parameterization of $(r,u,\theta)$.

\begin{figure*}
\epsscale{1.0} \plotone{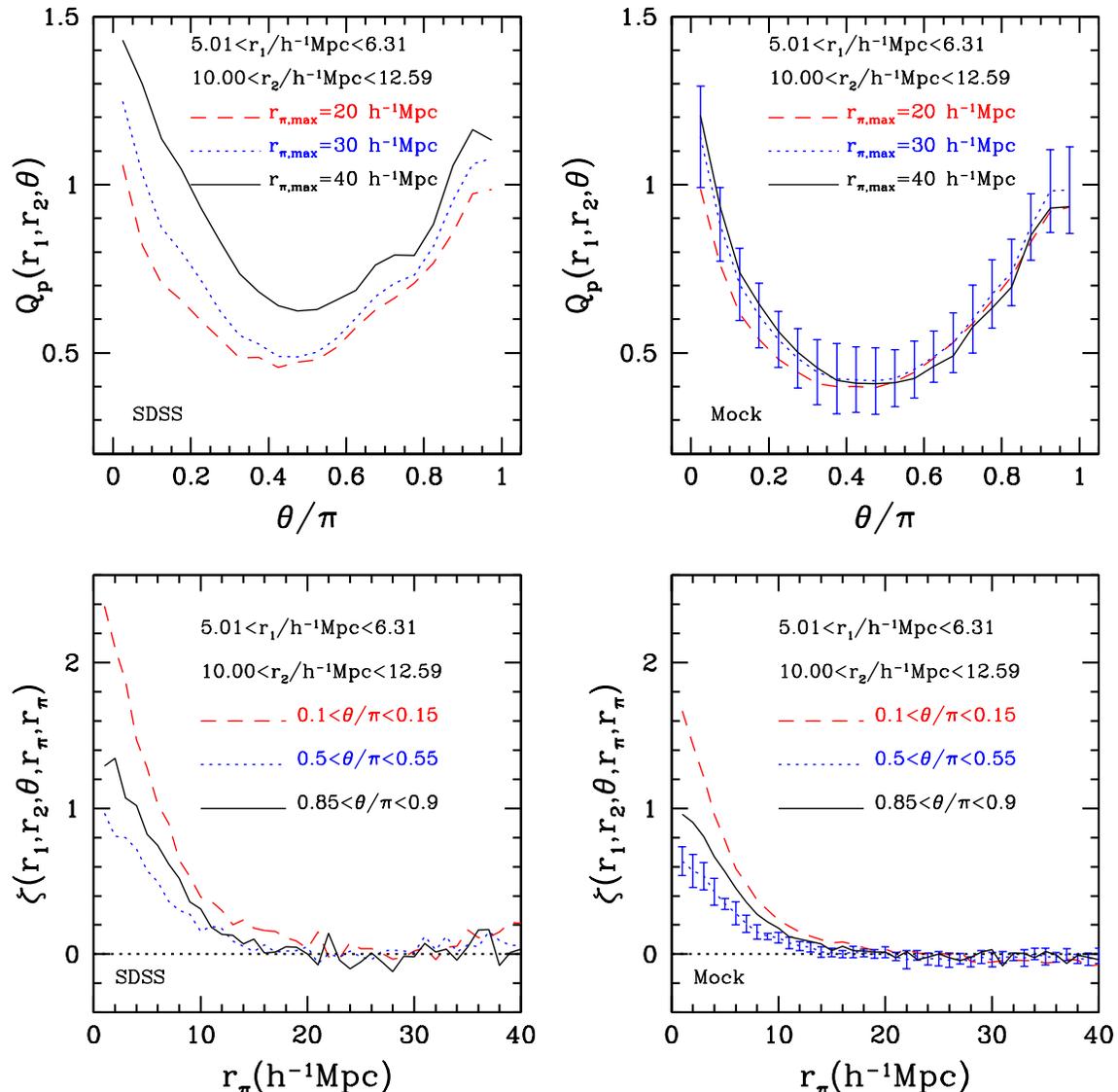} \caption{Top panels: measurements of
projected 3PCF, $Q_p(r_1,r_2,\theta)$, using $r_{\pi,\rm{max}}=20\mpchi$ (red
dashed lines), $30\mpchi$ (blue dotted lines), and $40\mpchi$ (black solid
lines) for the real data (left) and mocks (right). The measurements are made
for the stellar-mass subsample $10<\log M_s<10.5$, and in specific cases of
$r_1$ and $r_2$. The measurement errors for the mocks are only shown for one
sample for clarity. Bottom panels: data (left) and mock (right) measurements
of $\zeta(r_1,r_2,\theta,r_{\pi1},r_{\pi2})$ for the case of
$r_{\pi1}=r_{\pi2}$ in the same ranges of $r_1$ and $r_2$ but three different
ranges of $\theta$, denoted by different types of lines as labeled. (A color
version of this figure is available in the online journal.)}
\label{fig:pimax}
\end{figure*}
To find a suitable $r_{\pi,\rm{max}}$ for the integration of the 3PCF
$\zeta(r_1,r_2,\theta,r_{\pi1},r_{\pi2})$ along the line of sight, we
investigate its effect using the stellar-mass sample $\rm{M3}$, which is the
largest stellar-mass subsample. The measurements of projected 3PCF, $Q_p$,
using different $r_{\pi,\rm{max}}$ for the real data and mocks are shown in
the top left and top right panels of Figure \ref{fig:pimax}, respectively. In
the calculation of $Q_p$, the $r_{\pi,\rm{max}}$ for both $\Pi(r_1,r_2,r_3)$
and $w_p(r_p)$ are kept the same in order to test its effect. It is clear
that the amplitude and shape of $Q_p$ are dependent on the choice of
$r_{\pi,\rm{max}}$, consistent with the conclusions of \cite{McBride-11a}. We
find that the mocks provide reasonably good consistency with the real data
when $r_{\pi,\rm{max}}=20\mpchi$ is used. However, the $Q_p$ of the real data
is significantly larger than the mock predictions when a larger
$r_{\pi,\rm{max}}$ is employed.

To investigate this further, we show in the bottom panels of Figure
\ref{fig:pimax} the measurements of $\zeta(r_1,r_2,\theta,r_{\pi1},r_{\pi2})$
for the case of $r_{\pi1}=r_{\pi2}$ in specific ranges of $r_1$, $r_2$, and
$\theta$ as labeled in the figure. As shown in the real data, most of the
signals are included within $r_{\pi,\rm{max}}=20\mpchi$ and the measurements
beyond $r_{\pi,\rm{max}}=20\mpchi$ are mostly contributed to by the noise of
uncorrelated triplets. However, such noise is not well reproduced in the
mocks. So in order to have a reasonable estimate of the errors, as well as
suppressing the contribution from the noise, we set the maximum of
integration over the line of sight as $r_{\pi,\rm{max}}=20\mpchi$ for both
$r_{\pi12}$ and $r_{\pi23}$, the same as in \cite{McBride-11a}. Changing
$r_{\pi,\rm{max}}$ for the 2PCF from $40\mpchi$ to $20\mpchi$ will only have
effects at a few percent level. So we still use the measurements of the 2PCFs
from the integrations to $40\mpchi$ along the line of sight. The binning of
$r_{\pi}$ is defined as $\Delta r_{\pi}=1\mpchi$, the same as in
\cite{Jing-Borner-04a}. The choice of $\Delta r_{\pi}$ will only
significantly affect the resulting correlation functions when a very large
bin size is used \citep{McBride-11a}. The possible noise produced by null
random triplet counts in $r_{\pi}$ bins is reduced by requiring $\zeta=0$ in
such bins to ignore their contributions. The resulting projected 3PCF will
not be significantly affected when a large random sample is used.

\section{Results}\label{sec:results}
\subsection{Stellar-mass Dependence}
\begin{figure*}
\epsscale{1.0} \plotone{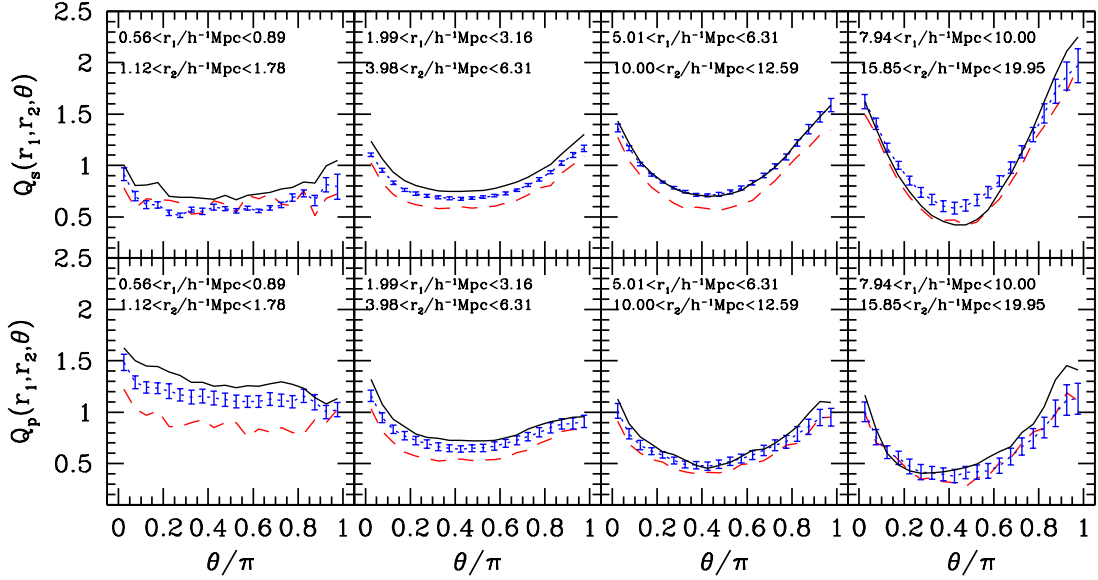} \caption{Reduced 3PCF in redshift (upper
panels) and projected spaces (bottom panels) at different scales and triangle
shapes. They are only shown for the case of $\bar{r}_2=2\bar{r}_1$. Lines of
different colors denote different luminosity samples, with the black solid,
blue dotted, and red dashed lines for $L2$, $L3$, and $L4$, respectively. The
errors are only shown for one sample for clarity. (A color version of this
figure is available in the online journal.)} \label{fig:qlum}
\end{figure*}
Before exploring the stellar-mass dependence of the 3PCF, we show in Figure
\ref{fig:qlum} the luminosity dependence of $Q_s$ in redshift space and $Q_p$
in projected space for comparison with other work
\citep[e.g.][]{McBride-11a}. For simplicity and without loss of generality,
we only show $Q_s$ and $Q_p$ for the configuration of
$\bar{r}_2/\bar{r}_1=2$. Lines of different colors denote different
luminosity samples, with the black solid, blue dotted, and red dashed lines
for $L2$, $L3$, and $L4$, respectively (from faint samples to luminous ones).
The sample $L1$ is not shown due to its small volume, as explained in Section
3.1. There is a clear luminosity dependence of $Q_s$ and $Q_p$, with more
luminous galaxies showing lower $Q$, consistent with the findings of
\cite{Jing-Borner-04a}, \cite{Gaztanaga-05} and \cite{McBride-11a}. Such
dependence is weaker on large scales where the errors on the measurements are
also large.

\begin{figure*}
\epsscale{1.0} \plotone{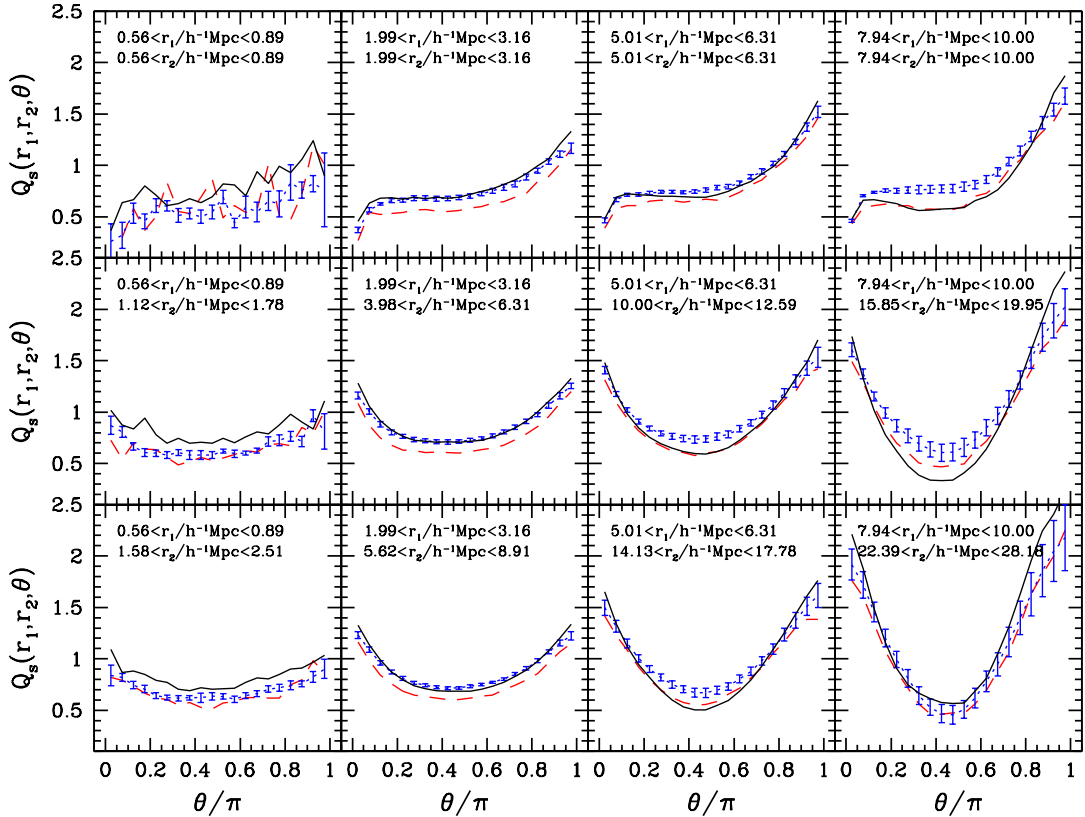} \caption{Redshift-space reduced 3PCF,
$Q_s$, for different stellar-mass samples at various scales and triangle
shapes. Lines of different colors denote different stellar samples, with the
black solid, blue dotted, and red dashed lines for $M2$, $M3$, and $M4$,
respectively. The errors are only shown for one sample for clarity. (A color
version of this figure is available in the online journal.)}
\label{fig:qstellar}
\end{figure*}
We show in Figure \ref{fig:qstellar} the stellar-mass dependence of the
redshift-space reduced 3PCF, $Q_s$. Lines of different colors denote samples
of different stellar mass, with the black solid, blue dotted, and red dashed
lines for $M2$, $M3$, and $M4$ (from low mass to high mass), respectively.
Different panels present the results of $Q_s$ for different scales and
shapes. The stellar-mass dependence of $Q_s$ is very weak, compared with the
luminosity dependence. The stellar-mass dependence mostly shows up on small
scales, where there is a trend that more massive galaxies have a lower
amplitude of $Q_s$. This is roughly consistent with the finding that more
luminous galaxies have a smaller $Q_s$ \citep{Jing-Borner-04a,McBride-11a},
considering the strong correlation of stellar mass and luminosity \cite[see
e.g., Figure 2 in][]{Li-06a}.

\begin{figure}
\epsscale{1.2} \plotone{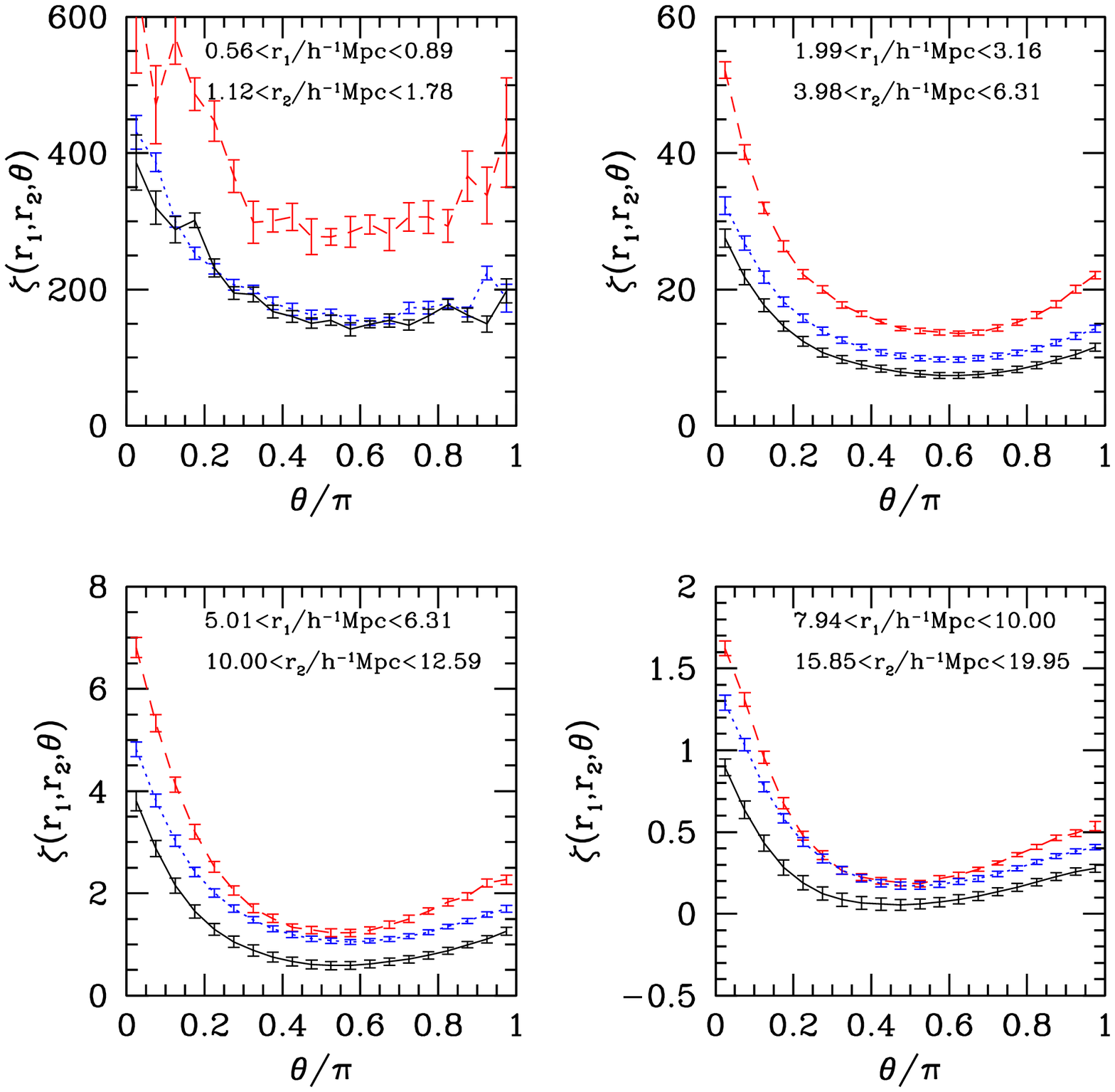} \caption{Redshift-space 3PCF, $\zeta(r_1,r_2,\theta)$,
for different stellar-mass samples at various scales. The lines are the same as in Figure
\ref{fig:qstellar}. Only the case of $\bar{r}_2=2\bar{r}_1$ is shown.
(A color version of this figure is available in the online journal.)}
\label{fig:zetas}
\end{figure}
A way to characterize the dependence of 3PCF on stellar mass is to compare
the redshift-space 3PCF, $\zeta(r_1,r_2,\theta)$, rather than the reduced
value, $Q_s$. We show in Figure \ref{fig:zetas} the stellar-mass dependence
of $\zeta(r_1,r_2,\theta)$ for the case of $\bar{r}_2=2\bar{r}_1$ at four
different scales. We find that more massive galaxies are clustered more
strongly than the less massive ones in the 3PCF, consistent with the results
of the 2PCF \citep{Li-06a}. Such a stellar-mass dependence of $\zeta$ is
contrary to the stellar-mass dependence of $Q_s$, because the stellar-mass
dependence of $\xi^2$ in Equation (\ref{eqn:qs}) is much stronger than that
of $\zeta$. This can be understood from the larger galaxy bias of those more
massive galaxies. In the second-order perturbation theory, the galaxy 3PCF
$\zeta(r_1,r_2,r_3)$ can be expressed as \citep[see
e.g.,][]{Fry-93,Pan-05,McBride-11b,Marin-13}
\begin{equation}
\zeta=b_1^3\zeta_m+b_2b_1^2(\xi_{m,1}\xi_{m,2}+\xi_{m,2}\xi_{m,3}+\xi_{m,3}\xi_{m,1}), \label{eqn:zetab}
\end{equation}
where $\zeta_m(r_1,r_2,r_3)$ is the dark-matter 3PCF, and $\xi_{m,i}(r_i)$ is
the dark-matter 2PCF at separation $r_i$. The parameters $b_1$ and $b_2$ are
the linear and nonlinear bias factors. To leading order, we can approximate
the galaxy 3PCF as $\zeta\sim b_1^3\zeta_m$. Since the galaxy 2PCF, $\xi$,
scales as $\xi\sim b_1^2\xi_m$, we conclude that $\xi^2$ varies much stronger
($\propto b_1^4$) with different stellar-mass samples than $\zeta$.

When assuming constant galaxy biases $b_1$ and $b_2$, the natural derivation
from Equation (\ref{eqn:zetab}) is \citep[see, e.g.,][]{Bernardeau-02}
\begin{equation} Q_g=\frac{Q_m}{b_1}+\frac{b_2}{b_1^2}, \label{eqn:bias} \end{equation}
where $Q_g$ and $Q_m$ are the galaxy and dark-matter reduced 3PCFs,
respectively. In the linear bias model where $b_2=0$, $Q_g$ is proportional
to $1/b_1$, consistent with the trend shown in Figure \ref{fig:qstellar}.
However, such explanations of the dependence of the reduced 3PCF, $Q_s$, on
the stellar mass are only qualitative, since the galaxy bias on these
nonlinear scales is strongly scale dependent, as shown in Figure
\ref{fig:2pcfbias}. Moreover, when using the redshift-space 3PCF to measure
the galaxy bias, the RSD effect should be carefully taken into account
\citep{Scoccimarro-99,Pan-05}. We will explore the stellar-mass dependence of
the 3PCF in more detail using the halo-occupation models in a companion
paper.

In the left panels of Figure \ref{fig:qstellar}, we find that $Q_s$ shows a
weaker shape dependence on small scales ($<2\mpchi$), regardless of the
stellar mass. \cite{McBride-11a} measured $Q_s$ on scales from $3\mpchi$ to
$27\mpchi$ and found significant configuration dependence, consistent with
our measurements on similar scales. In order to identify the correct scale of
the triangle, we define the scale $r_{\rm max}$ as the maximum of
$(r_1,r_2,r_3)$. Our measurements on these small scales show that the
clustering hierarchy, $\zeta\propto\xi^2$, would be more accurate when
applied on small scales.

Comparing Figures \ref{fig:qstellar} and \ref{fig:zetas}, we find that while
$\zeta$ decreases with the scale, the scale dependence of $Q_s$ is twofold.
For elongated triangle shapes ($\theta{\sim}0$ and $\pi$), $Q_s$ increases
with scale, while for the perpendicular triangles ($\theta\sim\pi/2$), there
is a weak trend for $Q_s$ to decrease as the scale increases. These two
effects make $Q_s$ show a much stronger shape dependence on larger scales. On
large scales, galaxies tend to be more clustered in linear structures, such
as filaments \citep{Scoccimarro-01}. Less massive galaxies have a stronger
shape dependence and are thus possibly more abundant in these linear
structures. Since galaxies of different stellar mass generally have different
shapes of the 2PCF \citep{Li-06a}, it is indeed expected to find the
degeneracy of the stellar-mass and shape dependence on large scales.

The case $r_1=r_2$ for $Q_s$ is very special, because it shows a quite
different configuration dependence as in other panels. With such triangles,
the 3PCF can measure scales less than $r_1$ when $\theta<\pi/3$ and the
maximum scale $r_{\rm max}$ of the triangle is $r_1$. It seems that the
configuration dependence of $Q_s$ shown in this way is determined by $r_{\rm
max}$ of the triangle. If we define $r_{\rm max}$ as the scale, we can use
the angle $\theta_{\rm max}$ subtended by $r_{\rm max}$ as the configuration
$\theta$. Therefore, the mild shape dependence of $Q_s$ seen in the top
panels of Figure~\ref{fig:qstellar} when $\theta<\pi/3$ is caused by the fact
that the $\theta_{\rm max}$ subtended by $r_{\rm max}$ is only in the range
of $[\pi/3,\pi/2]$, where the variation of $Q_s$ with $\theta_{\rm max}$ is
small. So the shape dependence of $Q_s$ is correlated with, but not
determined by, its scale dependence. The downturn of signals in the first one
or two bins ($\theta{\sim}0$) in the top panels of Figure \ref{fig:qstellar}
can be caused by low S/N of small numbers of triplet counts or by the
residual fiber-collision effect since the fiber collision in our measurements
is not well corrected below the fiber-collision scale (${\sim}0.1\mpchi$).

\begin{figure*}
\epsscale{1.0} \plotone{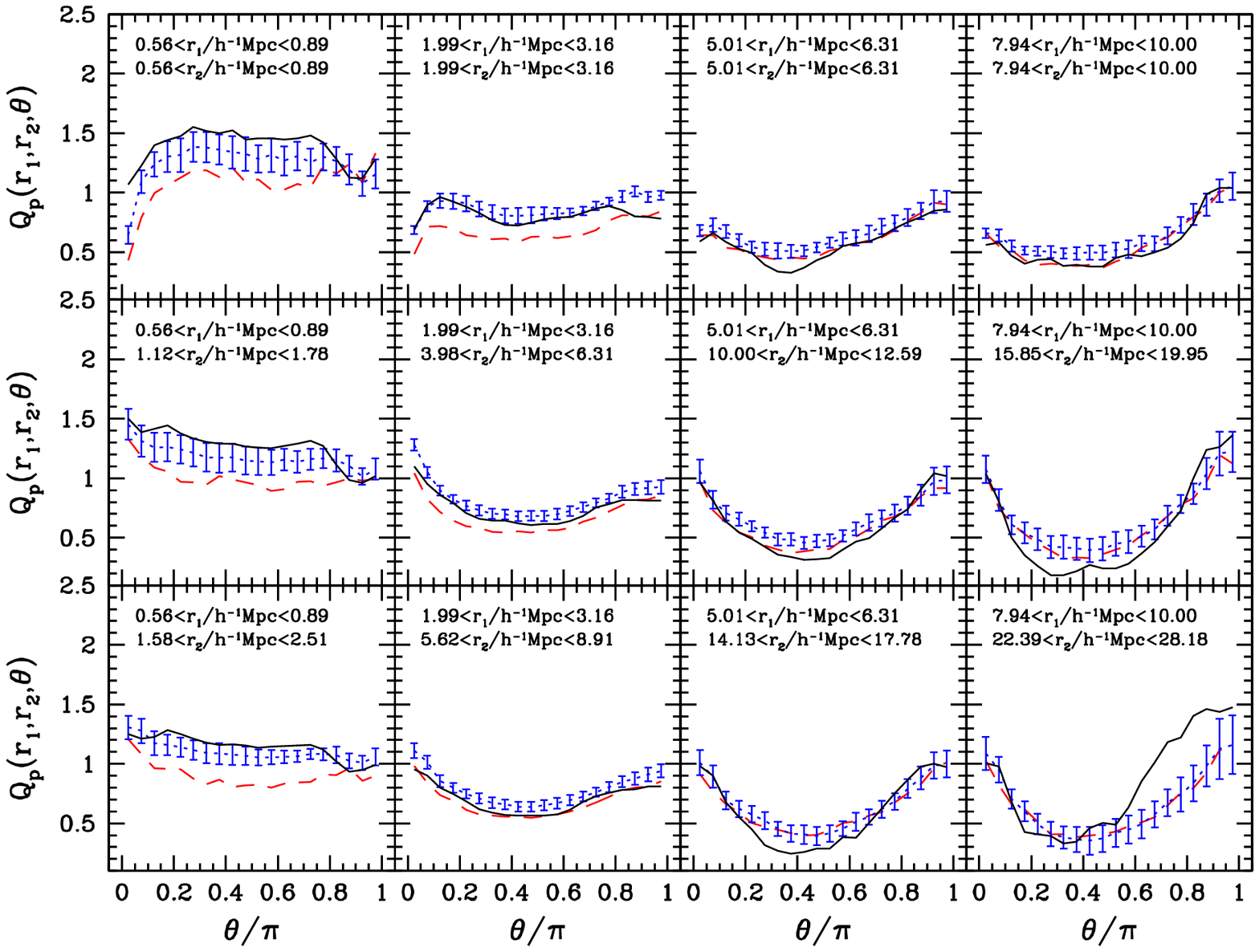} \caption{Same as in Figure
\ref{fig:qstellar}, but for the projected-space reduced 3PCF, $Q_p$. (A color
version of this figure is available in the online journal.)}
\label{fig:pstellar}
\end{figure*}
\begin{figure}
\epsscale{1.2} \plotone{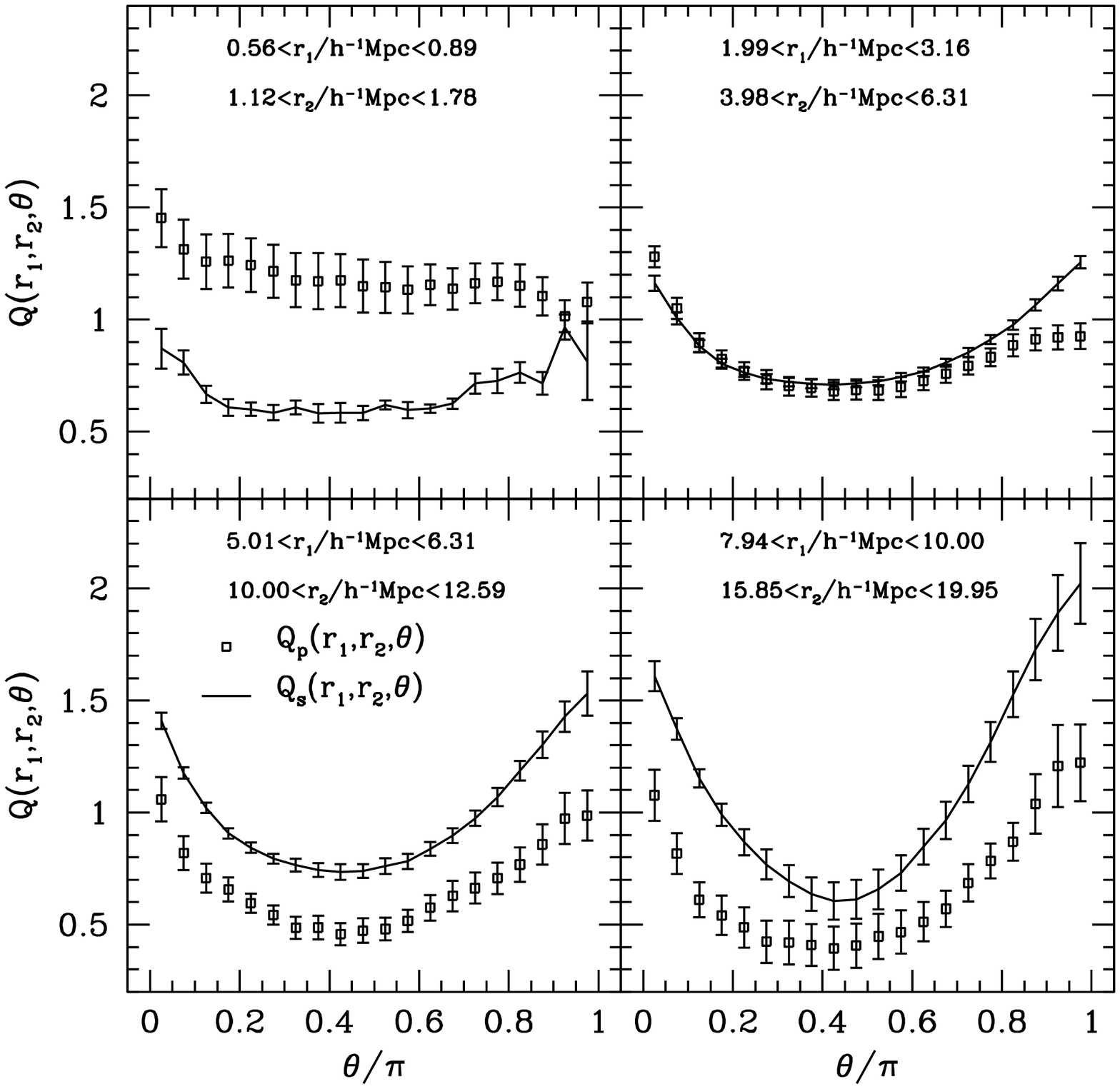} \caption{Comparison of the scale and shape dependence of $Q_s$ (redshift space) and
$Q_p$ (projected space) for the stellar-mass sample of $10.0<\log(M_s)<10.5$. The points are for $Q_p$ and lines for
$Q_s$.} \label{fig:pq}
\end{figure}
We present the stellar-mass dependence for the projected-space reduced 3PCF,
$Q_p(r_1,r_2,\theta)$, in Figure \ref{fig:pstellar}. Similar to the reduced
3PCF, $Q_s$, in redshift space, $Q_p$ does not have a strong stellar-mass
dependence, implying that the stellar-mass dependence of the reduced 3PCF is
not significantly affected by the RSD effect. In Figure \ref{fig:pq}, we
compare $Q_s$ and $Q_p$ for the sample of $10.0<\log(M_s)<10.5$. We find that
$Q_p$ generally has a weaker shape dependence than $Q_s$, which different
from the conclusion of \cite{McBride-11a}, who shows that the projected-space
$Q_p$ recovers more shape dependence that is lost in $Q_s$ due to RSD. It is
possible that the shape dependence shown in their figures is somewhat
smoothed out by their choice of the parameterization ($r_1,r_2,r_3$).
\cite{Marin-08} compared the measurements of the reduced 3PCF in real and
redshift spaces for scales down to  $r_1=1.5\mpchi$ using the $N$-body
simulations. They found a significant configuration dependence of $Q$ in real
space and concluded that the RSDs attenuate the shape dependence of $Q$.
Since the projected correlation function is directly related to the
real-space correlation function \citep{Jing-Borner-04a}, the weaker shape
dependence of $Q_p$ seems to contradict the predictions from the simulations
in real space. This could be caused by the possible selection effects or
systematics in the survey that suppress the shape dependence in the projected
space. We will study this further using mock galaxy catalogs in future work.

\subsection{Color Dependence}
\begin{figure*}
\epsscale{1.0} \plotone{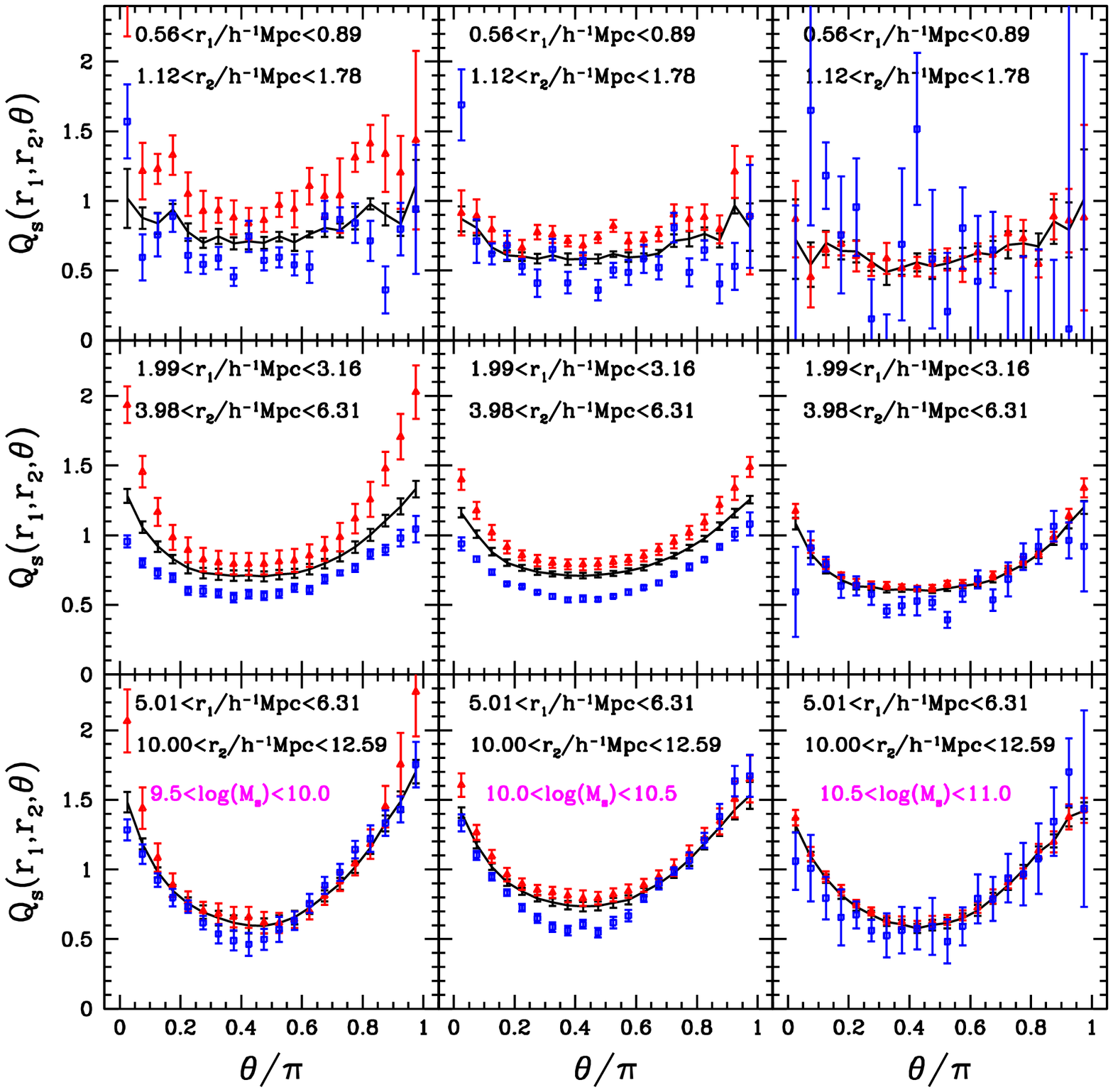} \caption{Redshift-space reduced 3PCF,
$Q_s$, for different color samples at different stellar mass and scales. The
left, middle, and right panels are for the samples of $9.5<\log M_s<10$,
$10<\log M_s<10.5$, and $10.5<\log M_s<11$, respectively. The red triangles
and blue squares are for red and blue galaxies defined in Section
\ref{sec:data}, respectively. The results for all the galaxies in the
stellar-mass samples are shown as black lines for comparison. We only display
the results for $\bar{r}_2/\bar{r}_1=2$. (A color version of this figure is
available in the online journal.)} \label{fig:qcolor}
\end{figure*}
\begin{figure*}
\epsscale{1.0} \plotone{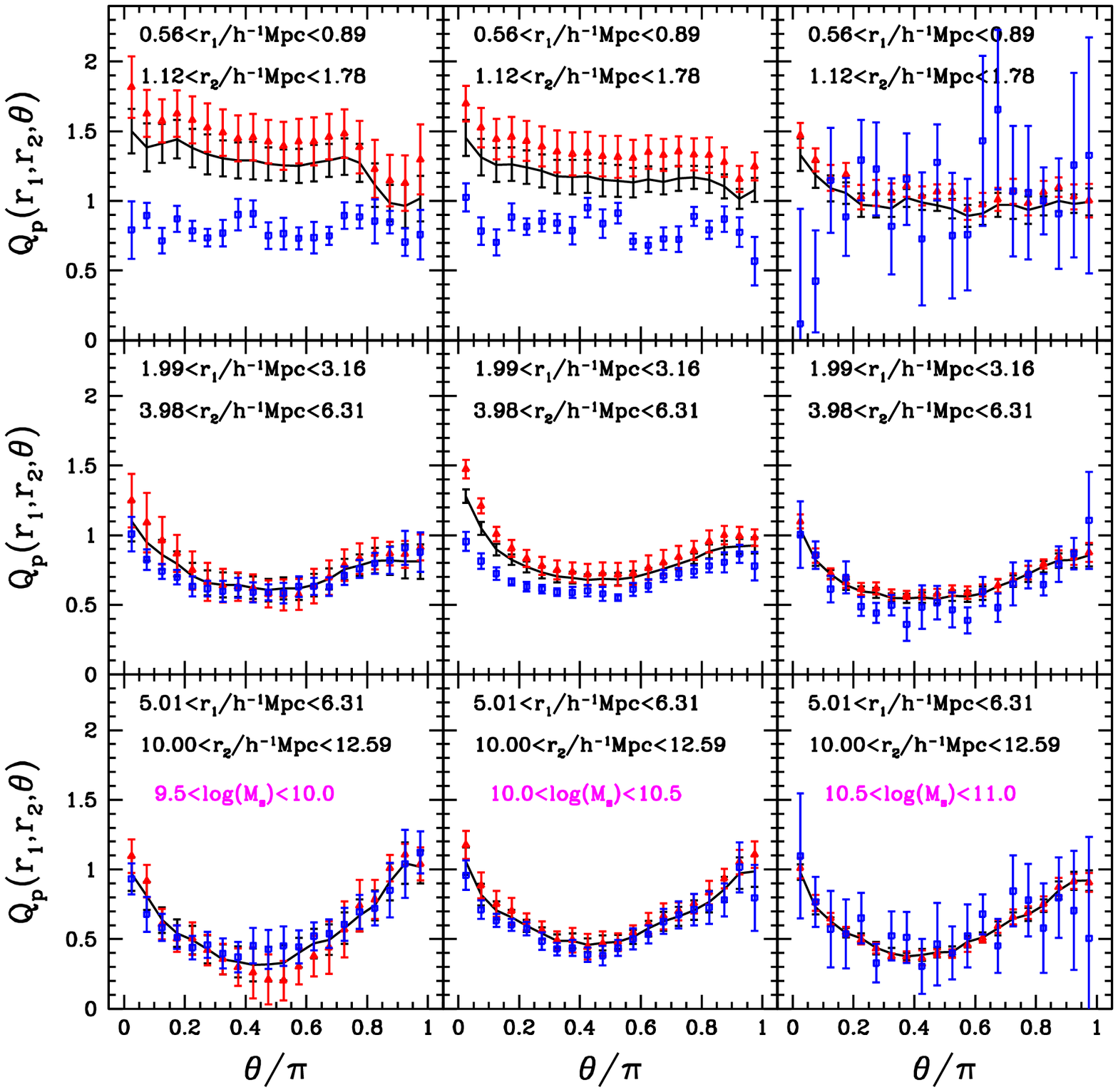} \caption{Same as in Figure
\ref{fig:qcolor}, but for the projected-space reduced 3PCF, $Q_p$. (A color
version of this figure is available in the online journal.)}
\label{fig:pcolor}
\end{figure*}
We show the color dependence of $Q_s$ and $Q_p$ for different stellar-mass
samples in Figures \ref{fig:qcolor} and \ref{fig:pcolor}. The red triangles
and blue squares are for the red and blue galaxies defined in Section
\ref{sec:data}, respectively. The results for all the galaxies in the
stellar-mass samples are also shown as black lines. We only display the
results for the case of $\bar{r}_2/\bar{r}_1=2$. The measurement errors for
the blue galaxies in the top right panels of Figures \ref{fig:qcolor} and
\ref{fig:pcolor} are very large because of the shot-noise errors of small
numbers of triplet counts at these small scales. The differences between the
3PCFs of the red and blue galaxies are more significant for less massive
samples and on smaller scales. For the most massive galaxy sample in the
right panels of Figures \ref{fig:qcolor} and \ref{fig:pcolor}, there is no
significant dependence on color, considering the large errors on the
measurements of blue galaxies. Although the fraction of red to blue galaxies
changes with the stellar-mass subsamples considered, it only affects the
shot-noise levels of the measurements of different color samples and should
not change the color dependence of $Q_s$ and $Q_p$.

Using high-resolution $N$-body simulations, \cite{Guo-Jing-09a} demonstrate
that Equation (\ref{eqn:bias}) can be applied to measure the linear and
nonlinear galaxy bias even on small scales where the second-order
perturbation theory fails and the bias factors become scale dependent. The
red galaxies are found to have a higher clustering amplitude than the blue
galaxies in the 2PCF measurements \citep{Zehavi-05,Li-06a}, i.e., red
galaxies have a larger linear galaxy bias, $b_1$. For linear-bias models, we
expect $Q_s$ and $Q_p$ for red galaxies to be somewhat smaller than those for
blue galaxies, contrary to the results in Figures \ref{fig:qcolor} and
\ref{fig:pcolor}. This implies that the linear-bias model is not enough to
describe the galaxy bias of different color samples, i.e., the nonlinear bias
factor, $b_2$, would not be negligible and plays an important role in the
modeling of the reduced 3PCF for galaxies of different colors. Considering
that $b_1$ and $Q$ are both larger for red galaxies, we infer from Equation
(\ref{eqn:bias}) that the nonlinear bias $b_2$ of the red galaxies would also
be larger than that of the blue ones. Because of the scale dependence of the
bias factors, constant bias models are only accurate on large scales and
precise measurements of the dark matter 3PCF is required for accurate
estimation of the bias factors \citep{Guo-Jing-09b,Pollack-13}. Comparing
Figures \ref{fig:qcolor} and \ref{fig:pcolor}, it seems that the RSD effect
is stronger for the red galaxies, especially on small scales, while the blue
galaxies suffer less from the RSD.

In the 2PCF, the differences between the clustering strength of the red and
blue galaxies become smaller in more luminous galaxy samples and on larger
scales \citep{Li-06a}. We also find a similar trend in $Q_s$ and $Q_p$. In
the bottom panels of Figures \ref{fig:qcolor} and \ref{fig:pcolor}, $Q_s$ and
$Q_p$ for the red and blue galaxies are more similar on large scales and in
more massive galaxy samples. It implies that the galaxy bias factors would
have less dependence on color for massive galaxies on large scales, which
provides the opportunity to use massive galaxies irrespective of color to
correctly measure the large-scale bias in galaxy redshift surveys.

Considering the significant differences of red and blue galaxies in the
low-mass end, the 3PCF can provide powerful constraints to the formation and
evolution of the low mass galaxies. The higher reduced 3PCF $Q_s$ or $Q_p$
for these low-mass red galaxies on small scales reflects the different
distributions of red and blue galaxies within the dark matter halos. From the
perspective of the halo model \citep[see e.g.,][]{Fosalba-05}, the
small-scale contribution to the 3PCF mostly comes from the one-halo term
where triplets of galaxies reside in the same halos. Since the red galaxies
of such low stellar masses are mostly satellite galaxies in relatively
massive halos and most of these blue galaxies are central galaxies
\citep{Li-07, Zehavi-11}, the lack of one-halo central--satellite--satellite
and satellite--satellite--satellite contributions from the blue galaxies
results in the lower amplitudes of $Q_s$ and $Q_p$. Similarly, the small
differences of the reduced 3PCF for red and blue galaxies in the
high-stellar-mass samples reflect the similar satellite fractions of
different colors. The clustering differences in the low-mass end could also
relate to the different environments of the low-mass red and blue galaxies,
since the red galaxies tend to populate denser regions \citep{Hogg-03}. We
note that for low-mass samples, red galaxies have a stronger shape dependence
than the blue galaxies, especially on large scales. This indicates that the
low-mass red galaxies are more influenced by the large-scale structure and
preferentially populate the filamentary structures.

\section{Conclusions and discussions}
\label{sec:discussion}

In this paper, we measure the stellar mass and color dependence of the galaxy
3PCF using the SDSS DR7 main sample galaxies in the redshift range of
$0.001<z<0.5$. We also investigate the scale and shape dependence of the 3PCF
for samples of different stellar mass and color in both redshift and
projected spaces.

In the redshift space, the dependence of the reduced 3PCF, $Q_s$, on the
stellar mass is very weak. Slightly stronger dependence on stellar mass shows
up on small scales, where there is a trend that more massive galaxies have
lower amplitudes of $Q_s$, consistent with the finding that more luminous
galaxies have smaller $Q_s$ \citep{Jing-Borner-04a,McBride-11a}. Such a
behavior of $Q_s$ can be qualitatively understood from the larger linear bias
of more massive galaxies using the second-order perturbation theories. The
reduced 3PCF, $Q_s$, is also dependent on the scale and shape of the
triangle, which provides measurements of the 3D spatial distribution in the
galaxies. We find that the shape dependence of $Q_s$ is stronger on larger
scales, reflecting the fact that the filamentary structures dominate the
large-scale distribution. More massive galaxies show weaker shape dependence
of $Q_s$, which is possibly resulting from the high occupancy of low-mass
galaxies in the filamentary structures. The reduced 3PCF, $Q_s$, only has
weak shape dependence on small scales, regardless of the stellar mass,
meaning that the hierarchical clustering model $\zeta\propto\xi^2$ would be
more accurate on these small scales.

To study the effect of RSD, we also measure the similar dependence of the
projected-space reduced 3PCF $Q_p$ on stellar mass, scale, and the triangle
shape. We find no strong dependence of $Q_p$ on the stellar mass at different
scales and triangle shapes, as in the redshift space. It means that the weak
stellar-mass dependence is not caused by the RSD effect. Weaker shape
dependence is found for $Q_p$ than that of redshift-space 3PCF $Q_s$, which
is different from the expectation that RSD attenuates the shape dependence of
the reduced 3PCF. We will explore this effect in more detail using mock
galaxy catalogs in future work.

We also investigate the color dependence of $Q_s$ and $Q_p$ for different
stellar mass samples. In redshift space, the color dependence is stronger for
low-mass galaxies and on small scales, reflecting the different distributions
of low-mass red and blue galaxies in the small-scale structures. The low-mass
red galaxies have higher 2PCF clustering amplitudes and also higher $Q_s$
values, which indicates that the linear bias models are not sufficient to
predict the 3PCF distribution of various color samples. The higher amplitude
of $Q_s$ for red galaxies implies that the red galaxies reside in denser
environments than the blue galaxies. Red galaxies display stronger shape
dependence than blue galaxies in the low-mass samples on large scales, which
means that the low-mass red galaxies tend to populate the filamentary
structures. The high-mass red and blue galaxies generally show very similar
amplitudes and shape dependence of $Q_s$ on all scales probed, indicating
their similar spatial distributions.

In the projected space, the similar differences between the $Q_p$ for red and
blue galaxies are still significant for low-mass galaxies. But such
differences are smaller compared to those in the redshift space, especially
for the red galaxies. This is an indication of stronger RSD effects for the
low-mass red galaxy distribution. The behavior of the high-mass red and blue
galaxies are still similar as in the redshift space.

From the conclusions above, we can have a clearer picture of how the 3PCF
helps understand the galaxy spatial distribution and their formation and
evolution. For example, from the reduced 3PCF of the low-mass red galaxies,
we conclude that on large scales these galaxies preferentially populate the
filamentary structures. On small scales, they are more clustered than the
blue galaxies of similar masses and populate denser regions. The failure of
the linear bias model for the color samples is also consistent with the fact
that most of these low-mass red galaxies are satellite galaxies and their
nonlinear evolution is important.

Since the different behaviors in the spatial clustering of 3PCF reflect the
different formation and merger history of galaxies, we will use the 3PCF
measurements in this work to test the galaxy formation and evolution models
using detailed halo occupation models and also connect the relation of galaxy
stellar masses and host halo masses in our future work. The study of the
stellar mass dependence of the 3PCF will help further constrain the galaxy
formation and evolution models when combined with the information of the star
formation rates, merger histories, and even the active galactic nucleus
feedback of the different stellar mass samples in the semi-analytical models.

\acknowledgments We thank the anonymous referee for the helpful suggestions
that significantly improve the paper. This work is supported by NSFC (no.
11173045, no. 11233005, no. 11320101002, no. 11325314) and the CAS/SAFEA
International Partnership Program for Creative Research Teams (KJCX2-YW-T23).
HG was supported by NSF grant AST-0907947. CL acknowledges the support of the
100 Talents Program of the Chinese Academy of Sciences (CAS), Shanghai
Pujiang Program (no. 11PJ1411600) and the exchange program between the Max
Planck Society and the CAS.

Funding for the SDSS and SDSS-II has been provided by the Alfred P. Sloan
Foundation, the Participating Institutions, the National Science Foundation,
the U.S. Department of Energy, the National Aeronautics and Space
Administration, the Japanese Monbukagakusho, the Max Planck Society, and the
Higher Education Funding Council for England. The SDSS Web Site is
http://www.sdss.org/.

The SDSS is managed by the Astrophysical Research Consortium for the
Participating Institutions. The Participating Institutions are the American
Museum of Natural History, Astrophysical Institute Potsdam, University of
Basel, University of Cambridge, Case Western Reserve University, University
of Chicago, Drexel University, Fermilab, the Institute for Advanced Study,
the Japan Participation Group, Johns Hopkins University, the Joint Institute
for Nuclear Astrophysics, the Kavli Institute for Particle Astrophysics and
Cosmology, the Korean Scientist Group, the Chinese Academy of Sciences
(LAMOST), Los Alamos National Laboratory, the Max-Planck-Institute for
Astronomy (MPIA), the Max-Planck-Institute for Astrophysics (MPA), New Mexico
State University, Ohio State University, University of Pittsburgh, University
of Portsmouth, Princeton University, the United States Naval Observatory, and
the University of Washington.

\end{document}